# Collider Implications of Multiple Non-Universal Extra Dimensions

R. Ghavri,[a,b*] C.D. McMullen,[c†] and S. Nandi[a‡]

[a] *Department of Physics, Oklahoma State University*
*Stillwater, OK 74078, USA*

[b] *Department of Chemistry and Physics, Northwestern State University*
*Natchitoches, LA 71457, USA*

[c] *Department of Physics, Louisiana School for Math, Science, and the Arts*
*Natchitoches, LA 71457, USA*

**Abstract**

We consider multiple $\text{TeV}^{-1}$-size extra compact dimensions in an asymmetric string compactification scenario in which the SM gauge bosons can propagate into the $\text{TeV}^{-1}$-size extra dimensions while the SM fermions are confined to the usual SM $D_3$-brane. We calculate the contributions that the KK excitations of the gluons, $g^*$'s, make to the multijet cross sections in proton-proton collisions at the LHC energy. At very high $p_T$, the dijet signal will either be enhanced significantly due to virtual $g^*$ exchanges or place a lower bound on the compactification scale of about 8 TeV. We find that the dijet signal is very sensitive to three parameters – the compactification scale, the string scale, and the number of extra dimensions. Thus, although the dijet signal is much more sensitive to KK effects, the dijet signal alone does not provide sufficient information to deduce the number of extra dimensions nor the compactification scale. However, the three-jet signal, which is not sensitive to the string scale, can be analyzed in conjunction with the dijet signal to extract all three parameters. For proton-proton collisions at the LHC energy, the three-jet signal can be significantly enhanced by KK contributions for a compactification scale of about 4-5 TeV.

---

[*] email: ghavrir@nsula.edu
[†] email: cmcmullen@lsmsa.edu
[‡] email: s.nandi@okstate.edu

# 1. Introduction

Two broad classes of papers on the collider phenomenology of superstring-inspired [1] large extra compact dimensions include papers in the spirit of the original Arkani-Hamed, Dimopoulos, and Dvali (ADD) scenario [2], in which only the gravitons propagate into the bulk [3], and models where one or more of the standard model (SM) fields can also propagate into one or more extra compact dimensions. In the latter case, two subclasses include the universal model [4-6], in which all of the SM fields can propagate into one or more extra compact dimensions, and the non-universal model [7-12], in which the SM gauge fields can propagate into one or more extra compact dimensions while the SM fermions are confined to the usual $D_3$-brane. Much of the published research involving detailed calculations in models in which one or more SM fields can propagate into extra compact dimensions has been done for a single extra compact dimension – partly because the most stringent bounds arise from a single extra compact dimension, and partly because the number of parameters and complexity of the calculations increases with increasing number of extra compact dimensions. However, models in which the SM fields can propagate into multiple extra compact dimensions can lead to interesting collider phenomenology. Examples include dilepton production at the LHC [12], analyses of various compactification schemes [12-13], and bounds set by the Fermi constant [14].

In this work, we consider the non-universal model, in which the gluons can propagate into multiple TeV$^{-1}$-size extra compact dimensions while the fermions are confined to the SM $D_3$-brane. More specifically, we examine how the multijet production in high-energy hadronic colliders such as the Large Hadron Collider (LHC) depends on three parameters: the string scale $M_S$, and the number $\delta$ and size $1/\mu$ of the large extra compact dimensions. We calculate the contributions that the Kaluza-Klein (KK) excitations of the gluons, $g^*$'s, make to the production of multijet final states which arise from the direct production and exchanges of KK excitations of the gluons. We find that the number of extra compact dimensions does have a significant effect on the production of $g^*$'s in proton-proton collisions at the LHC energy. For example, with a $p_T$ cut of 2 TeV, the KK contribution to the total dijet cross section will be comparable to the SM dijet cross section for compactification scales up to 7 TeV if there is just one TeV$^{-1}$-size extra dimension, while if there are two TeV$^{-1}$-size extra dimensions the KK contribution to the total dijet cross section will be comparable to the SM dijet cross section for compactification scales up to 11-14 TeV if the string scale $M_S$ is 4-10 times larger than the compactification scale $\mu$ (i.e. $4 \leq \Lambda \leq 10$ where $\Lambda = M_S/\mu$).

Our paper is organized as follows. We present our formalism in Section 2, supplemented by additional details in the Appendix. The contributions that virtual $g^*$ exchanges make to dijet production are discussed in Section 3, while the contributions that single on-shell $g^*$'s make to three-jet production are presented in Section 4. We draw our conclusions in Section 5.

## 2. Formalism

In the non-universal model, the SM fermions are constrained to lie in the SM $D_3$-brane while the gauge fields can propagate into TeV$^{-1}$-size extra compact dimensions. There may be additional much higher-scale extra dimensions as in the asymmetric scenario [7]. Here we present the Feynman rules in the effective 4D theory needed to evaluate multi-jet cross sections in proton-proton collisions at the LHC energy.

Straightforward generalization of the 4D SM Lagrangian density leads to the $(4+\delta)$-D Lagrangian density,

$$\mathcal{L}_\delta = -\frac{1}{4} F^{MNa} F^a_{MN} + i\bar{q}\gamma^\mu D_\mu q \delta(y_1)\delta(y_2)\cdots\delta(y_\delta) \tag{1}$$

where $F^{MNa} = \partial^M A^{Na} - \partial^N A^{Ma} - g_{4+\delta} f^{abc} A^{Mb} A^{Nc}$ are the $(4+\delta)$-D gluon field strength tensors, $g_{4+\delta}$ is the $(4+\delta)$-D strong coupling, $A^{Ma}$ is the $(4+\delta)$-D gluon field, $\{a,b,c\}$ are the usual gluon color indices, $D_\mu$ is the usual 4D covariant derivative, $\{\mu,\nu\}$ are the usual 4-D spacetime indices, $\{M,N\} \in \{0,1,...,3+\delta\}$ are $(4+\delta)$-D spacetime indices, and the product of delta functions represents that the SM fermions are localized to the SM $D_3$-brane with $y_1 = y_2 = \cdots = y_\delta = 0$. For $\delta = 2$ the 6D Lagrangian density is

$$\mathcal{L}_6 = -\frac{1}{4} F^{\mu\nu a} F^a_{\mu\nu} - \frac{1}{2} F^{\mu 4a} F^a_{\mu 4} - \frac{1}{2} F^{\mu 5a} F^a_{\mu 5} + i\bar{q}\gamma^\mu D_\mu q \delta(y_1)\delta(y_2) \tag{2}$$

where the gauge choice $A^{4a} = A^{5a} = 0$ has been imposed [6,15]. As in Ref. [4], we consider compactification on an $(S_1 \times S_1 / Z_2)^2$ orbifold, corresponding to a 2D torus cut in half along $y_1$. That is, $0 \leq \varphi_1 \leq \pi$ and $-\pi \leq \varphi_2 \leq \pi$. The fields $A^{\mu a}(x,y)$ can then be Fourier expanded in terms of the compactified extra dimensions $y_{1,2} = r\,\varphi_{1,2}$ (assuming that the TeV-scale extra dimensions are symmetric – i.e. they have the same radius $r$) as

$$A^{\mu a}(x,y) = \frac{1}{\sqrt{2}\pi r}\left[ A^{\mu a}_{0,0}(x) + \sqrt{2} \sum_{n_1,n_2}^{\infty} A^{\mu a}_{n_1,n_2}(x)\cos(n_1\varphi_1 + n_2\varphi_2)\Xi(n_1 + n_2) \right] \tag{3}$$

where the modified step function $\Xi(n_1 + n_2) = 1$ if $n_1 + n_2 \geq 1$ or $n_1 = -n_2 \geq 1$ and 0 otherwise. The $\sqrt{2}$ reflects the rescaling of the gauge fields necessary to canonically normalize the kinetic energy terms [9,16]. The summation limits and modified step function reflect the compactification scheme ($0 \leq \varphi_1 \leq \pi$ and $-\pi \leq \varphi_2 \leq \pi$).

Integration over the compactified dimensions $y_1$ and $y_2$ then gives the effective 4D Lagrangian density. The masses of the KK excitations of the gluons arise from the integration of $F^{\mu 4a} F_{\mu 4}{}^a + F^{\mu 5a} F_{\mu 5}{}^a$ over $y_1$ and $y_2$:

$$-\frac{1}{2} \int_{y_1=0}^{\pi r} \int_{y_2=-\pi r}^{\pi r} \left( F^{\mu 4 a} F_{\mu 4}^a + F^{\mu 5 a} F_{\mu 5}^a \right) dy_1 dy_2$$

$$= -\frac{1}{4\pi^2 r^2} \int_{y_1=0}^{\pi r} \int_{y_2=-\pi r}^{\pi r} \left( \partial^4 A^{\mu a}(x,y) \partial_4 A_\mu^a(x,y) + \partial^5 A^{\mu a}(x,y) \partial_5 A_\mu^a(x,y) \right) dy_1 dy_2 \quad (4)$$

$$= -\frac{1}{2} \frac{n_1^2 + n_2^2}{r^2} \sum_{n_1,n_2}^{\infty} A_{n_1,n_2}^{\mu a}(x) A_{\mu a}^{n_1,n_2}(x) \Xi(n_1 + n_2)$$

The masses of the $g^*$'s are identified as

$$m_{(n_1,n_2)} = \mu \sqrt{n_1^2 + n_2^2} \quad (5)$$

where $\mu$ is the compactification scale ($\mu = 1/r$). The lowest-lying KK excitations up to $\vec{n} = 3$ include $g_{1,0}^*$, $g_{0,1}^*$, $g_{1,1}^*$, $g_{1,-1}^*$, $g_{2,0}^*$, $g_{0,2}^*$, $g_{2,1}^*$, $g_{2,-1}^*$, $g_{2,2}^*$, $g_{2,-2}^*$, $g_{3,0}^*$, and $g_{0,3}^*$.

The Feynman rules for the effective 4D couplings are derived in the Appendix. The results are tabulated in Fig. 1. KK number conservation for triple and quartic gluonic vertices involves an important subtlety. Consider, as an illustration, $g_{m_1,n_1}^*$ coupling to $g_{m_2,n_2}^*$ and $g_{m_3,n_3}^*$: KK number is conserved if $m_3 = m_1 + m_2$ and $n_3 = n_1 + n_2$, but is not conserved if $m_3 = m_1 + m_2$ and $n_1 = n_2 + n_3$, for example. However, a single $g^*$ can couple to a quark pair with a $\sqrt{2}$ enhancement compared to the SM owing to the delta function in the 6D Lagrangian density that confines the fermions to the SM D$_3$-brane. In this case, the SM D$_3$-brane absorbs the unbalanced four-momentum.

The $g^*$ propagator in the unitary gauge is

$$-i\Delta_{\mu\nu\vec{n}}^{ab} = -i\delta^{ab} \frac{g_{\mu\nu} - \frac{p_\mu p_\nu}{m_{\vec{n}}^2}}{p^2 - m_{\vec{n}}^2 + i m_{\vec{n}} \Gamma_{\vec{n}}} \quad (6)$$

where the tree-level decay width of the $g^*$ is $\Gamma_{n_1,n_2} = 2\alpha_S(Q) m_{n_1,n_2}$. The mass of the $g^*$ also enters into the computations when external $g^*$'s are present via summation over polarization states:

$$\sum \epsilon_{\mu\vec{n}}^{a*}(k,\sigma) \epsilon_{\nu\vec{n}}^{b}(k,\sigma) = \left( -g_{\mu\nu} + \frac{k_\mu k_\nu}{m_{\vec{n}}^2} \right) \delta^{ab} \quad (7)$$

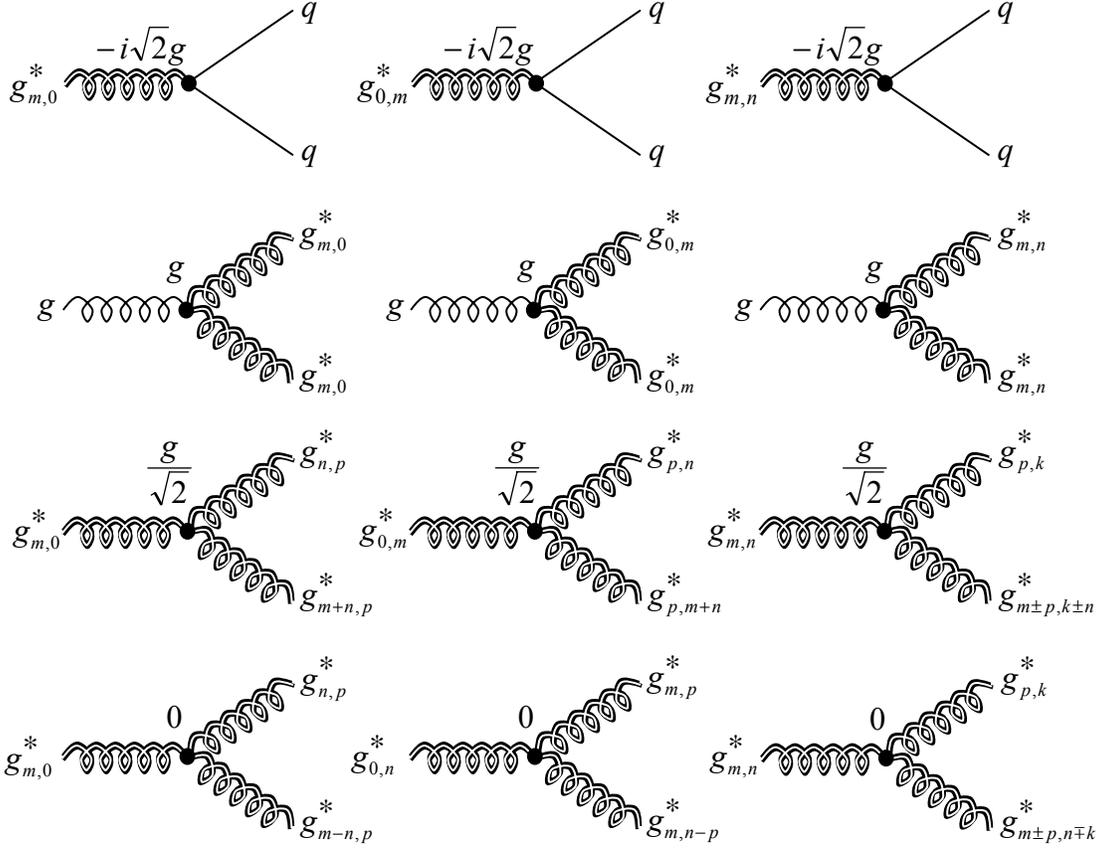

**Fig. 1**. Relative coupling strengths of vertices involving $g^*$'s. Only the overall factors are shown: The $q$-$\bar{q}$-$g^*$ vertex also involves the SU(3) matrix element and the Dirac $\gamma_\mu$ matrix, and triple gluonic vertices also include the usual SU(3) structure functions and the momenta factors. In $g^*_{m,n}$, $m$ must be positive, while $n$ may be negative if $m+n \geq 1$ or $m = -n \geq 1$. The zeroes indicate vertex factors that are not allowed.

### 3. Dijet Production

The dijet[§] cross section is enhanced by the exchange of virtual $g^*$'s in subprocesses with two initial quarks and two final-state quarks:

$q_i q_i \rightarrow q_i q_i$      $q_i q_j \rightarrow q_i q_j$      $q_i \bar{q}_i \rightarrow q_i \bar{q}_i$
$q_i \bar{q}_i \rightarrow q_j \bar{q}_j$      $q_i \bar{q}_j \rightarrow q_i \bar{q}_j$

The amplitudes for these subprocesses are the same as in the SM except for the replacement of the gluon propagator by a tower of $g^*$ propagators:

---

[§] We neglect the contributions from (2+N)-jet production where only two jets pass the experimental cuts.

$$D(p^2) = \frac{1}{p^2} + \sqrt{2} \sum_{n_1,n_2}^{\infty} \frac{1}{p^2 - m_{\tilde{n}}^2 + im_{\tilde{n}}\Gamma_{\tilde{n}}} \Xi(n_1 + n_2) \tag{8}$$

Thus, the amplitude-squared contains terms of the form

$$\frac{1}{2}[D^*(\hat{v})D(\hat{w}) + D(\hat{v})D^*(\hat{w})] = \sum_{\tilde{m},\tilde{n}}^{\infty} c_{\tilde{m}} c_{\tilde{n}} \frac{\tilde{v}'_{\tilde{m}}\hat{w}'_{\tilde{n}} + m_{\tilde{m}}\Gamma_{\tilde{m}}m_{\tilde{n}}\Gamma_{\tilde{n}}}{(\tilde{v}'^2_{\tilde{m}} + m_{\tilde{m}}^2\Gamma_{\tilde{m}}^2)(\hat{w}'^2_{\tilde{n}} + m_{\tilde{n}}^2\Gamma_{\tilde{n}}^2)} \tag{9}$$

where $\hat{v}$ and $\hat{w}$ are any of the three usual subprocess Mandelstam variables (i.e. $\hat{v}, \hat{w} \in \{\hat{s}, \hat{t}, \hat{u}\}$), $\tilde{v}'_{\tilde{n}}$ represents the subtraction of $m_{\tilde{n}}^2$ from $\hat{v}$ (i.e. $\tilde{v}'_{\tilde{n}} \equiv \hat{v} - m_{\tilde{n}}^2$), and the coefficients $c_{\tilde{n}}$ are defined according to $c_{(0,0)} = 1$ and $c_{\tilde{n} \neq 0} = \sqrt{2}$. This sum diverges logarithmically for $\delta = 2$, and even faster for $\delta \geq 2$. In practice, the sum is truncated when the $g^*$ mass reaches the string scale [10,17]. Therefore, the dijet signal is very sensitive not only to the compactification scale and $\delta$ but also to the cutoff scale. For $\delta = 2$, the divergence is logarithmic:

$$\sum_{n_1,n_2}^{\Lambda} \frac{1}{n_1^2 + n_2^2} \Xi(n_1 + n_1) \sim \frac{3\Lambda \ln(\Lambda)}{\Lambda - 1} \tag{10}$$

where $\Lambda$ represents the cutoff (related to the string scale $M_S$ by $M_S = \Lambda\mu$). For $\delta \geq 3$, the multi-sum diverges as a power law.

The total dijet cross section $\sigma(pp \to 2 \text{ jets})$ is given by

$$\sigma(pp \to 2 \text{ jets}) = \sum_{ab \to cd} \int_{\tau = 4p_T^2/s}^{1} \frac{d\mathcal{L}}{d\tau} \hat{\sigma}(ab \to cd) d\tau \tag{11}$$

where $\hat{\sigma}(ab \to cd)$ is the subprocess cross section and $d\mathcal{L}/d\tau$ is the parton luminosity:

$$\frac{d\mathcal{L}}{d\tau} = \int_{x_A = \tau}^{1} f_{a/A}(x_A, Q) f_{b/B}(x_B, Q) \frac{dx_A}{x_A} \tag{12}$$

Here, $f_{a/A}(x_A, Q)$ represents a parton distribution function evaluated at energy $Q$, $x$ is the momentum fraction, and $p_T$ is the transverse momentum. We employ the CTEQ distribution functions [18] in the parton luminosity evaluated at $Q = p_T$. We restrict the rapidity $y$ to lie within the range $|y| \leq 2.5$ and the transverse momentum $p_T$ to lie above $p_T^{\min}$.

We compute the KK signal and SM background at the tree-level. Although the relative uncertainty in the dijet cross section can be quite high, say 40%, at the tree-level due to the dependence on the arbitrary (at tree-level) parameter $Q$ and other factors, such as the choice of parton distributions, these uncertainties should cancel somewhat in the ratio of the similar calculations of the KK signal to the SM background. However, since this ratio cannot be measured directly, in order to be sure that a signal for new physics significantly stands out above the inherent uncertainties, we look for a KK contribution comparable to the SM background in which ~200 events per year are predicted at a proton-proton collider running at the LHC energy.

We denote by $\sigma_{KK}$ the KK contribution to the total dijet cross section $\sigma$: $\sigma = \sigma_{SM} + \sigma_{KK}$. This contribution, $\sigma_{KK}$, is illustrated in Fig.'s 2-3 for proton-proton collisions at the LHC energy for $\delta = 2$ and $\Lambda = 10$. The effect is actually quite large: For example, for a $p_T$ cut of 2 TeV, the KK contribution exceeds the SM contribution for compactification scales up to 13 TeV. This is much larger than the KK effect for $\delta = 1$ [11] where the KK contribution exceeds the SM contribution for compactification scales up to 7 TeV, and is enhanced even further for higher values of $\delta$, as shown in Fig. 4. Thus, the dijet cross section is very sensitive to the number of large extra dimensions. Furthermore, Fig. 5 illustrates that the dijet cross section is also very sensitive to the cutoff scale $\Lambda = M_S / \mu$ for $\delta \geq 2$, owing to the divergent form of the sum in the effective propagator for the virtual $g^*$ exchanges.

The total dijet cross section will be significantly enhanced if there is at least one non-universal extra dimension with a compactification scale less than 7 TeV. However, the KK dijet cross section is sensitive to three parameters – the string scale $M_S$, and the size $1/\mu$ and number $\delta$ of extra dimensions. Thus, if a proton-proton collider at the LHC energy does not observe a significant enhancement to the SM dijet cross section at very high $p_T$, then a minimum bound may be placed on the compactification scale based on the case of one extra dimension (because one extra dimension yields the smallest KK effects). However, if a proton-proton collider at the LHC energy observes a new physics enhancement to the SM dijet cross section, the total dijet production rate itself is not enough to deduce the number, structure, or size of the extra dimensions, or the string scale. For example, with a $p_T$ cut of 2 TeV, a total dijet cross section on the order of 0.1 pb could be caused by a single non-universal extra dimension with a compactification scale of 7 TeV, but it could also be caused by two non-universal extra dimensions with a compactification scale of 13 TeV and a string scale of 130 TeV or two non-universal extra dimension with a compactification scale of 10 TeV and a string scale of 40 TeV, etc. This ambiguity can be removed by analyzing additional signals. For example, three-jet production is not sensitive to the string scale. Dilepton production [12] also shows promise for helping to differentiate among compactification schemes.

Intuitively, it might seem beneficial to look for the resonant production of the just the lightest KK gluons – i.e. $g^*_{1,0}$, $g^*_{0,1}$, $g^*_{1,1}$, and $g^*_{1,-1}$ – instead of comparing the entire dijet signal to the SM background. However, in Ref. [11], in the context of a single extra non-universal dimension, the dijet differential cross section $\frac{d\sigma}{dm}$ did not provide a good signal when plotted as a function of the invariant mass $m$ of the $g^*$, which subsequently decays into a $q\bar{q}$ pair. One reason for this is that the $g^*$ has a large decay width, $\Gamma_{n_1,n_2} = 2\alpha_S(Q) m_{n_1,n_2}$, so the signal is not as tall nor sharp as it is in many other resonant productions. Also, the decay of the $g^*$ results in two high-$p_T$ jets, meaning that the KK signal will only be significant compared to the SM background for very high-$p_T$ cuts, which in turn severely limit the total number of anticipated events. The net result is that a greater KK enhancement is expected for the total dijet cross section at a proton-proton collider running at the LHC energy.

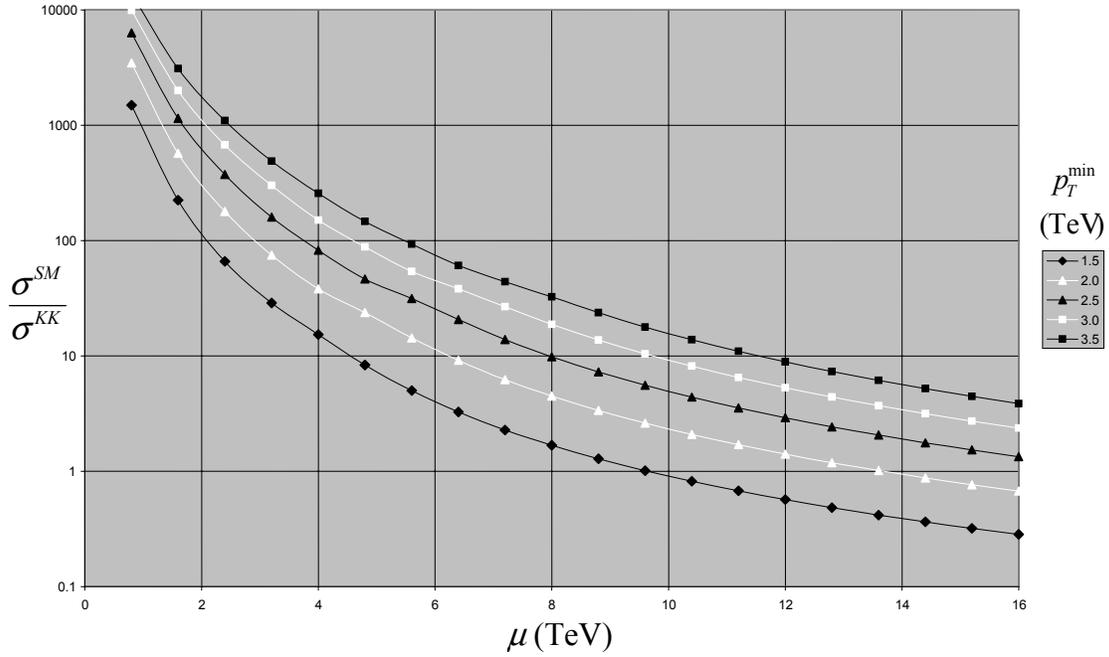

**Fig. 2**. The ratio of the KK contribution $\sigma_{KK} = \sigma - \sigma_{SM}$ toward the total dijet cross section $\sigma$ to the SM dijet cross section $\sigma_{SM}$ at the LHC is illustrated as a function of the compactification scale $\mu$ for various transverse momentum cuts $p_T^{\min}$ with $\delta = 2$ and $\Lambda = 10$. (For $p_T^{\min} = 3.5\,\text{TeV}$, $\sigma_{KK}$ is less than 0.001 pb for $\mu > 11\,\text{TeV}$.)

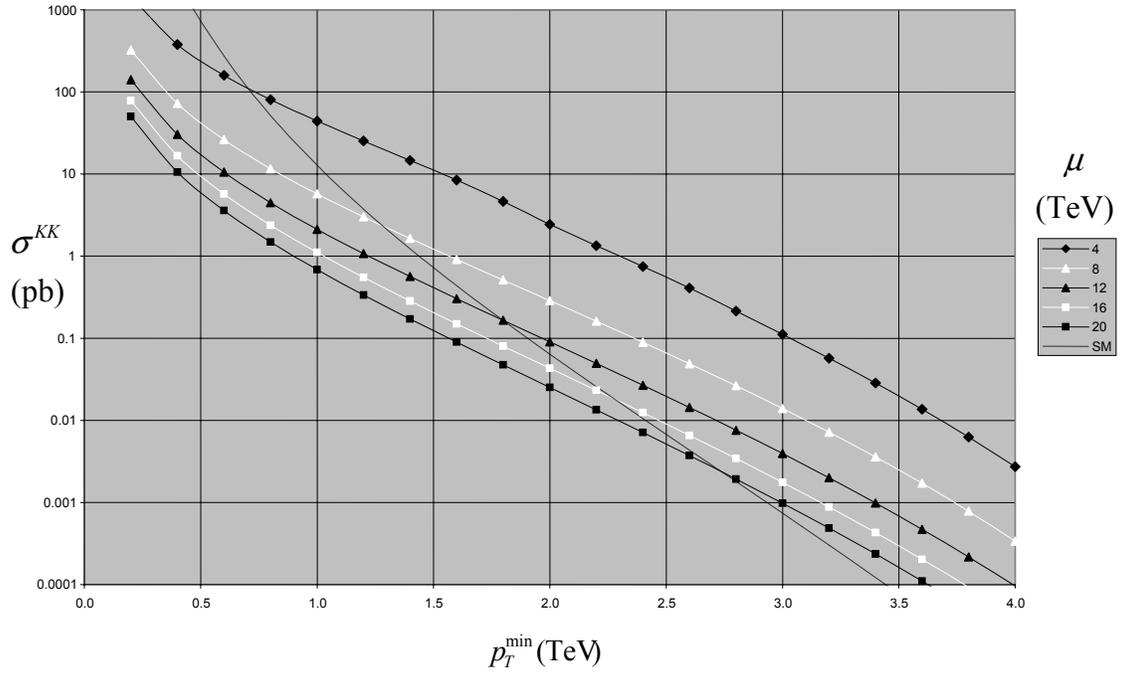

**Fig. 3**. The KK contribution $\sigma_{KK} = \sigma - \sigma_{SM}$ toward the total dijet cross section $\sigma$ at the LHC is illustrated as a function of the transverse momentum cut $p_T^{\min}$ for various compactification scales $\mu$ with $\delta = 2$ and $\Lambda = 10$.

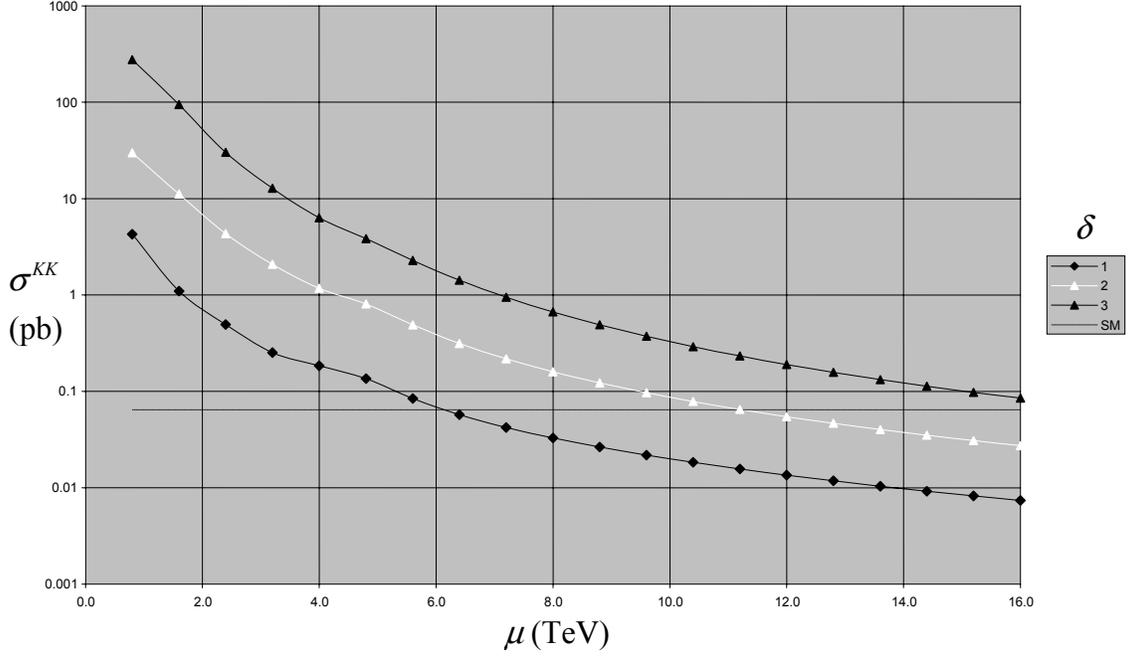

**Fig. 4**. The KK contribution $\sigma_{KK} = \sigma - \sigma_{SM}$ toward the total dijet cross section $\sigma$ at the LHC is illustrated as a function of the compactification scale $\mu$ for different numbers of extra non-universal dimensions $\delta$ with $p_T^{\min} = 2\,\text{TeV}$ and $\Lambda = 4$ for $\delta \geq 2$ (no cutoff $\Lambda$ is imposed for $\delta = 1$ because the propagator sum converges rapidly).

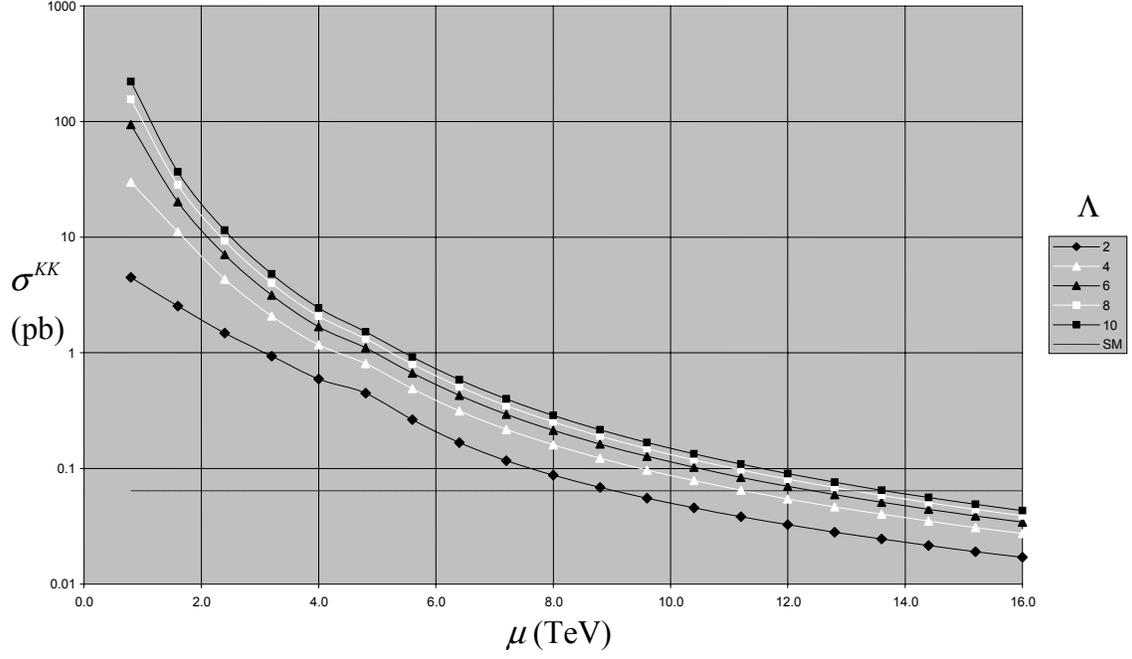

**Fig. 5**. The the KK contribution $\sigma_{KK} = \sigma - \sigma_{SM}$ toward the total dijet cross section $\sigma$ at the LHC is illustrated as a function of the compactification scale $\mu$ for various cutoffs $\Lambda = M_S / \mu$ with $\delta = 2$ and $p_T^{\min} = 2\,\text{TeV}$.

**4. Three-jet Production**

The production of one jet and an on-shell $g^*$ that decays to two jets leads to a three-jet event. The subprocesses leading to the production of one jet and an on-shell $g^*$ include:

$$q\bar{q} \rightarrow gg^*_{m,n} \qquad qg \rightarrow qg^*_{m,n} \qquad \bar{q}g \rightarrow \bar{q}g^*_{m,n}$$

KK number conservation demands that any $g^*$ propagators have the same mode ($m,n$) as the external $g^*$ such that there is no summation over modes in these propagators (in contrast to the case of dijet production). Instead, the three-jet cross section involves a double summation over possible final state modes ($m,n$) – finite as restricted by the available collider energy.

The amplitudes-squared for these subprocesses have the same form as in Ref. [11]. The difference is that there is a greater degeneracy of KK states with increasing number of large extra dimensions. In the case of one extra dimension, the three-jet cross section is dominated by the first mode – the contribution of the $g^*_2$ is on the order of one percent. In the case of two or more extra dimensions, the three-jet cross section is dominated by the $\delta$ lowest-order KK modes (e.g. the $g^*_{1,0}$ and $g^*_{0,1}$), and the lowest-lying mixed modes (e.g. the $g^*_{1,1}$ and $g^*_{1,-1}$) also make a significant contribution. Thus, the three-jet cross section in $\delta$ extra dimensions is somewhat greater than $\delta$ times the three-jet cross section in one extra dimension.

In addition to the dijet cuts, for three jets we also constrain the azimuthal angle $\phi$ and pseudorapidity $\eta$ of the final-state jets to satisfy $R = \sqrt{(\Delta\phi)^2 + (\Delta\eta)^2} \geq 0.4$. We neglect the contributions to the three-jet cross section that arise from virtual $g^*$ exchanges when no external $g^*$'s are produced since these processes include an extra factor of $\alpha_S(Q)$. We employ FORM [19], a symbolic manipulation program, to compute the amplitude-squares for the KK signal, and compute the SM three-jet background according to the outline of Ref. [20] – all at the tree level. As in the case of dijet production, in order to be sure that a signal for new physics significantly stands out above the inherent uncertainties, we look for a KK contribution comparable to the SM background in which ~200 events per year are predicted at a proton-proton collider running at the LHC energy.

The three-jet cross section is illustrated in Fig.'s 6-7. Fig. 6 shows peaks at $p_T^{min} = \frac{k\mu}{2}$ where $k \in \{1,2,...\}$ corresponding to $g_{k,0}^*$ and $g_{0,k}^*$, and peaks at $\frac{k\mu\sqrt{2}}{2}$ corresponding to $m_{k,k} = k\mu\sqrt{2}$. The effect that the number of extra dimensions has on the three-jet cross section is shown in Fig. 8. In contrast to dijet production, the three-jet cross section is not sensitive to the string scale $M_S$: Whereas dijet production effectively involves a sum over KK excitations in the propagator, each three-jet process involves only a few Feynman diagrams (in order to conserve KK number at the tree-level). Although three-jet production does include a sum over processes, involving all of the KK modes, the primary contributions to the three-jet cross section arise from the lowest-lying KK modes (e.g., for $\delta = 2$, $g_{1,0}^*, g_{0,1}^*, g_{1,1}^*, g_{1,-1}^*$). If $\Lambda \geq 2$, then the value of $\Lambda$ has virtually no effect on the three-jet cross section. Thus, the three-jet cross section effectively depends only upon the number of extra dimensions $\delta$ and the compactification scale $\mu$.

The presence of two non-universal extra dimensions will have a significant effect on the three-jet cross section in proton-proton collisions at the LHC energy with a $p_T$ cut of 1 TeV if the compactification scale is on the order of 3 TeV or less. Two non-universal extra dimensions has a substantial effect compared to one extra dimension. For $\delta \geq 3$, each additional extra dimension stretches the bound on the compactification scale by about 0.2 TeV. Even higher $p_T$ cuts can stretch the bound up to another TeV.

**5. Conclusions**

In this work, we have investigated the phenomenology of a class of string-inspired models with multiple non-universal extra dimensions where the SM fermions are confined to the SM D$_3$-brane while the gluons can propagate into multiple TeV$^{-1}$-size extra dimensions. Specifically, we have calculated the effects that the KK excitations of the gluons have on multijet final states at proton-proton colliders at the LHC energy. We find that the KK excitations make a significant contribution to the total dijet cross section relative to the SM background for very high $p_T$ if there is at least one non-universal extra dimension and the compactification scale is less than 7 TeV. We also find that the dijet cross section is sensitive to the number, size, and structure of the extra dimensions, in addition to the string scale. For example, with a $p_T$ cut of 2 TeV, a proton-proton collider at the LHC energy could observe a significant enhancement to the total dijet cross section with a single non-universal extra dimension with a compactification scale up to 7 TeV, or with two non-universal extra dimensions with a compactification scale up to 13 TeV and a string scale up to 130 TeV, or two non-universal extra dimension with a compactification scale up to 10 TeV and a string scale up to 40 TeV, etc.

In contrast to dijet production, we find that three-jet production is sensitive to the number, size, and structure of the extra dimensions, but is effectively independent of the string scale. Two non-universal extra dimensions will have a significant effect on the three-jet cross section in proton-proton collisions at the LHC energy with a $p_T$ cut of 1 TeV if the compactification scale is on the order of 3 TeV or less. For three or more non-universal extra dimensions, each additional extra dimension stretches the bound on the compactification scale by about 0.2 TeV. The bound can be stretched up to another TeV with very high $p_T$ cuts.

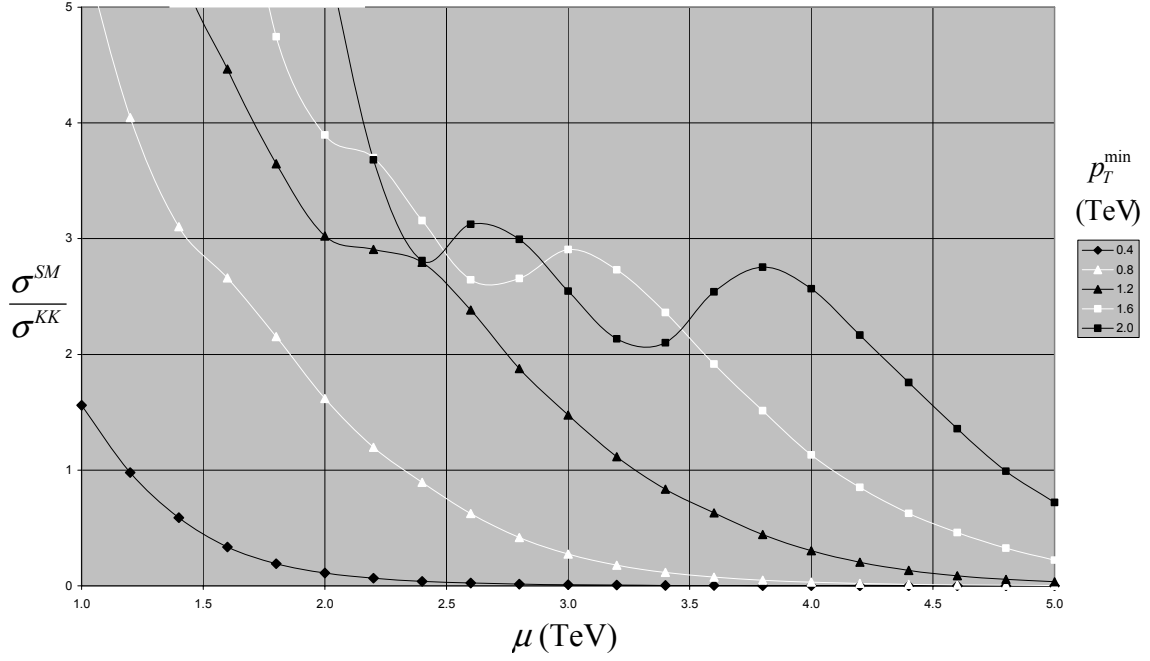

**Fig. 6**. The ratio of the KK contribution $\sigma_{KK} = \sigma - \sigma_{SM}$ toward the total three-jet cross section $\sigma$ to the SM three-jet cross section $\sigma_{SM}$ at the LHC is illustrated as a function of the compactification scale $\mu$ for various transverse momentum cuts $p_T^{min}$ with $\delta = 2$. (For $p_T^{min} = 1.2\,\text{TeV}$, $\sigma_{KK} < 10^{-3}$ pb for $\mu > 4.0\,\text{TeV}$; for $p_T^{min} \geq 1.6\,\text{TeV}$, $\sigma_{KK} < 10^{-3}$ pb.)

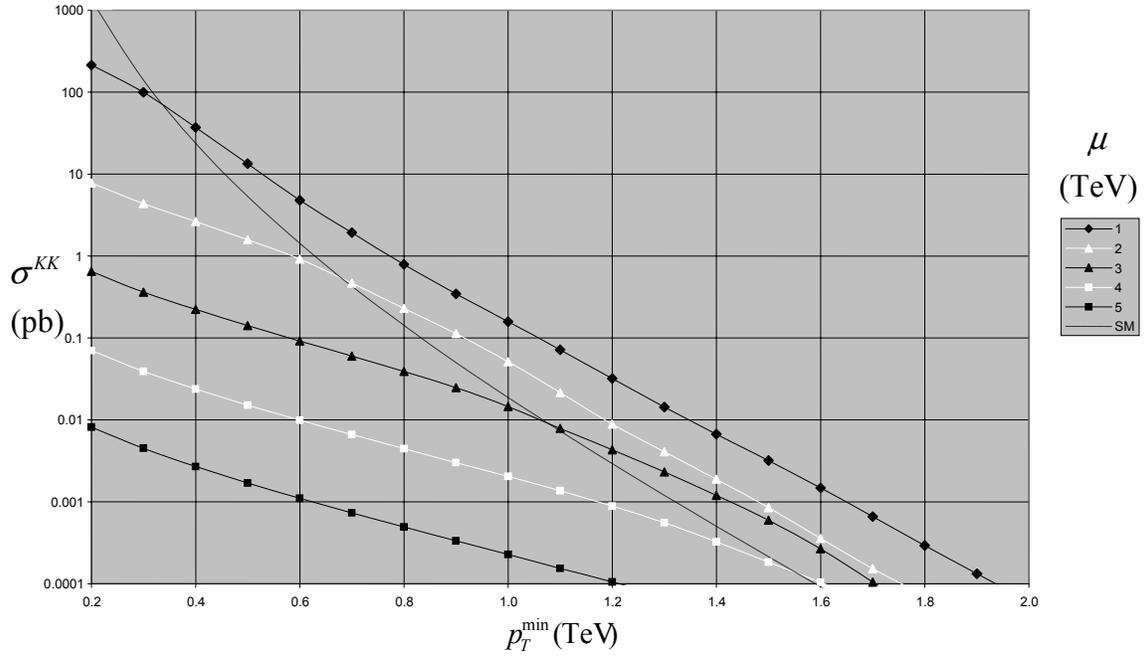

**Fig. 7**. The KK contribution $\sigma_{KK} = \sigma - \sigma_{SM}$ toward the total three-jet cross section $\sigma$ at the LHC is illustrated as a function of the transverse momentum cut $p_T^{min}$ for various compactification scales $\mu$ with $\delta = 2$.

If a proton-proton collider at the LHC energy does not observe a significant enhancement to the SM dijet cross section at very high $p_T$, then a minimum bound may be placed on the compactification scale based on the case of one extra dimension (because one extra dimension yields the smallest KK effects). However, if a proton-proton collider at the LHC energy observes a new physics enhancement to the SM dijet cross section, the total dijet production rate itself is not enough to deduce the number, structure, or size of the extra dimensions, or the string scale. This ambiguity can be removed by analyzing additional signals. For example, three-jet and four-jet production are not sensitive to the string scale.

The research of SN and RG is supported by the US Department of Energy Grant numbers DE-FG02-04ER41306 and DE-FG02-04ER46140. CM would like to thank OSU for its kind hospitality during this research.

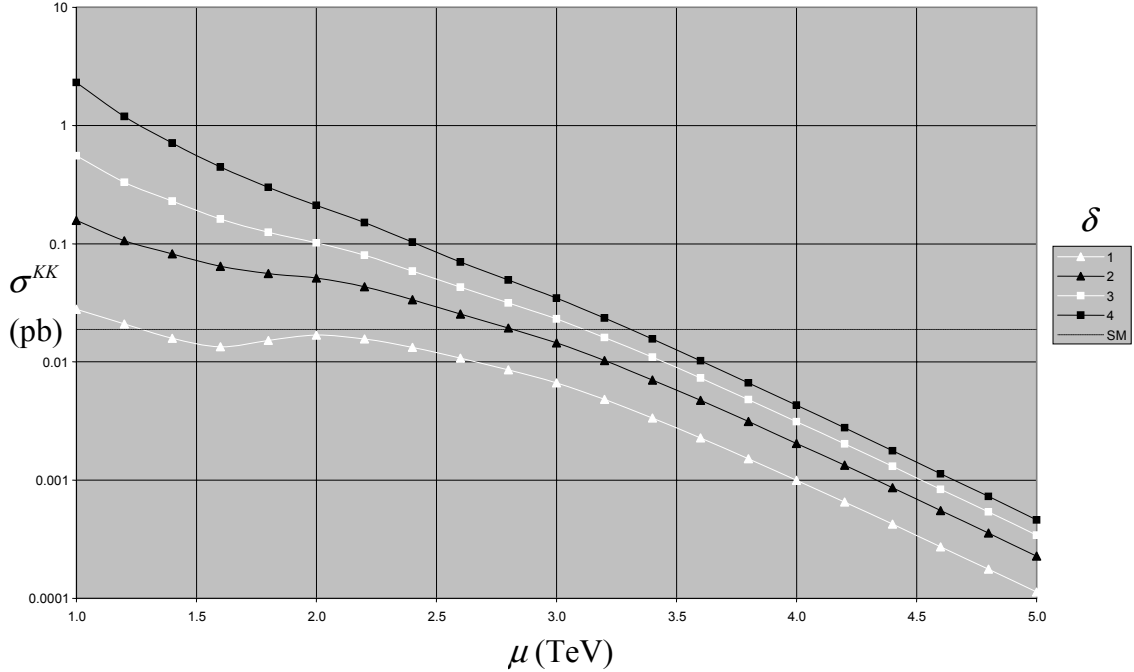

**Fig. 8**. The KK contribution $\sigma_{KK} = \sigma - \sigma_{SM}$ toward the total three-jet cross section $\sigma$ at the LHC is illustrated as a function of the compactification scale $\mu$ for different numbers of extra non-universal dimensions $\delta$ with $p_T^{\min} = 1\,\text{TeV}$.

**Appendix**

Here we present the Feynman rules in the effective 4D theory needed to evaluate multi-jet cross sections in proton-proton collisions at the LHC energy. For $\delta = 2$ the 6D Lagrangian density is

$$\mathcal{L}_6 = -\frac{1}{4} F^{\mu\nu a} F^a_{\mu\nu} - \frac{1}{2} F^{\mu 4a} F^a_{\mu 4} - \frac{1}{2} F^{\mu 5a} F^a_{\mu 5} + i\bar{q}\gamma^\mu D_\mu q \delta(y_1)\delta(y_2) \tag{A1}$$

where the gauge choice $A^{4a} = A^{5a} = 0$ has been imposed. We consider compactification on an $(S_1 \times S_1 / Z_2)^2$ orbifold, corresponding to a 2D torus cut in half along $y_1$. That is, $0 \leq \varphi_1 \leq \pi$ and $-\pi \leq \varphi_2 \leq \pi$. The fields $A^{\mu a}(x,y)$ can then be Fourier expanded in terms of the compactified extra dimensions $y_{1,2} = r\,\varphi_{1,2}$ (assuming that the TeV-scale extra dimensions are symmetric – i.e. they have the same radius $r$) as

$$A^{\mu a}(x,y) = \frac{1}{\sqrt{2}\pi r}\left[A^{\mu a}_{0,0}(x) + \sqrt{2}\sum_{n_1,n_1}^{\infty} A^{\mu a}_{n_1,n_2}(x)\cos(n_1\varphi_1 + n_2\varphi_2)\Xi(n_1+n_2)\right] \tag{A2}$$

where the modified step function $\Xi(n_1 + n_2) = 1$ if $n_1 + n_2 \geq 1$ or $n_1 = -n_2 \geq 1$ and 0 otherwise. The $\sqrt{2}$ reflects the rescaling of the gauge fields necessary to canonically normalize the kinetic energy terms. The summation limits and modified step function reflect the compactification scheme ($0 \leq \varphi_1 \leq \pi$ and $-\pi \leq \varphi_2 \leq \pi$).

Integration over the compactified dimensions $y_1$ and $y_2$ then gives the effective 4D Lagrangian density. The masses of the KK excitations of the gluons arise from the integration of $F^{\mu 4a} F^a_{\mu 4} + F^{\mu 5a} F^a_{\mu 5}$ over $y_1$ and $y_2$:

$$-\frac{1}{2}\int_{y_1=0}^{\pi r}\int_{y_2=-\pi r}^{\pi r}\left(F^{\mu 4a}F^a_{\mu 4} + F^{\mu 5a}F^a_{\mu 5}\right)dy_1 dy_2$$

$$= -\frac{1}{4\pi^2 r^2}\int_{y_1=0}^{\pi r}\int_{y_2=-\pi r}^{\pi r}\left(\partial^4 A^{\mu a}(x,y)\partial_4 A^a_\mu(x,y) + \partial^5 A^{\mu a}(x,y)\partial_5 A^a_\mu(x,y)\right)dy_1 dy_2 \tag{A3}$$

$$= -\frac{1}{2}\frac{n_1^2 + n_2^2}{r^2}\sum_{n_1,n_2}^{\infty} A^{\mu a}_{n_1,n_2}(x) A^{n_1,n_2}_{\mu a}(x)\Xi(n_1+n_2)$$

The masses of the $g^*$'s are identified as

$$m_{(n_1,n_2)} = \mu\sqrt{n_1^2 + n_2^2} \tag{A4}$$

where $\mu$ is the compactification scale ($\mu = 1/r$).

The Feynman rules for vertices involving $g^*$'s follow from the interaction terms in the effective 4D Lagrangian density. The $q$-$q$-$g^*$ interaction term in the effective 4D Lagrangian density is

$$-g_6 \int_{y_1=0}^{\pi r} \int_{y_2=-\pi r}^{\pi r} \bar{q}(x)\Gamma^\mu T^a A_\mu^a(x,y)q(x)\delta(y_1)\delta(y_2)dy_1 dy_2$$

$$= -g\bar{q}(x)\gamma^\mu T^a q(x)\left[A_{\mu 0}^a(x) + \sqrt{2}\sum_{n_1,n_2}^{\infty} A_{\mu a}^{n_1,n_2}(x)\Xi(n_1+n_2)\right]$$

(A5)

where the 4D strong coupling constant $g$ is related to $g_6$ by $g = g_6/\sqrt{2}\pi r$. Thus, the $q$-$\bar{q}$-$g^*$ vertex receives a factor of $\sqrt{2}$ enhancement, compared to the SM $q$-$\bar{q}$-$g$ vertex. The cubic interaction terms in the effective 4D Lagrangian density are

$$-i\frac{g_6}{2}f^{abc}\int_{y_1=0}^{\pi r}\int_{y_2=-\pi r}^{\pi r} A_\mu^b(x,y)A_\nu^b(x,y)\left[\partial^\mu A^{\nu a}(x,y) - \partial^\nu A^{\mu a}(x,y)\right]dy_1 dy_2$$

$$= -i\frac{g}{2}f^{abc}\Biggl\{A_{\mu b}^{(0,0)}A_{\nu c}^{(0,0)}\left(\partial^\mu A_{\nu a}^{(0,0)} - \partial^\nu A_{\mu a}^{(0,0)}\right) + 3A_{\mu b}^{(0,0)}\sum_{m,n}A_{\nu c}^{(m,n)}\left(\partial^\mu A_{\nu a}^{(m,n)} - \partial^\nu A_{\mu a}^{(m,n)}\right)$$

$$+ \frac{1}{\sqrt{2}}\sum_{\substack{m_1,n_1,p_1\\m_2,n_2,p_2}}A_{\mu b}^{(m_1,m_2)}A_{\nu c}^{(n_1,n_2)}\left(\partial^\mu A_{\nu a}^{(p_1,p_2)} - \partial^\nu A_{\mu a}^{(p_1,p_2)}\right)\delta(m_1,n_1,p_1,m_2,n_2,p_2)$$

$$+ \frac{3}{\sqrt{2}}\sum_{\substack{m_1,n_1\\p_1,n_2}}A_{\mu b}^{(m_1,0)}A_{\nu c}^{(n_1,n_2)}\left(\partial^\mu A_{\nu a}^{(p_1,n_2)} - \partial^\nu A_{\mu a}^{(p_1,n_2)}\right)\left(\delta_{m_1+n_1,p_1} + \delta_{m_1+p_1,n_1}\right)$$

$$+ \frac{3}{\sqrt{2}}\sum_{\substack{n_1,m_2\\n_2,p_2}}A_{\mu b}^{(0,m_2)}A_{\nu c}^{(n_1,n_2)}\left(\partial^\mu A_{\nu a}^{(n_1,p_2)} - \partial^\nu A_{\mu a}^{(n_1,p_2)}\right)\left(\delta_{m_2+n_2,p_2} + \delta_{m_2+p_2,n_2}\right)\Biggr\}$$

(A6)

where $\delta(m_1,n_1,p_1,m_2,n_2,p_2) \equiv \delta_{m_1+n_1,p_1}\delta_{m_2+n_2,p_2} + \delta_{m_1+p_1,n_1}\delta_{m_2+p_2,n_2} + \delta_{n_1+p_1,m_1}\delta_{n_2+p_2,m_2}$. The relative coupling strengths are summarized in Fig. 1.